\documentclass[showpacs,twocolumn, prb]{revtex4-1}
\usepackage{amsmath}
\usepackage{graphicx}
\usepackage{dcolumn}
\usepackage{bm}

\begin{document}

\title{Zone-Boundary Phonon Induced Mini Band Gap Formation in Graphene }
\author{B.S. Kandemir and A. Mogulkoc}
\date{\today }

\begin{abstract}
We investigate the effect of electron- $\mathrm{A}_{1g}$ phonon coupling on
the gapless electronic band dispersion of the pristine graphene. The
electron-phonon interaction is introduced through a Kekul\'{e}-type
distortion giving rise to inter-valley scattering between $\mathrm{K}$ and $%
\mathrm{K}^{\prime }$ points in graphene. We develop a Fr\"{o}lich type
Hamiltonian within the continuum model in the long wave length limit. By
presenting a fully theoretical analysis, we show that the interaction of
charge carriers with the highest frequency zone-boundary phonon mode of $%
\mathrm{A}_{1g}$-symmetry induces a mini band gap at the corners of the
two-dimensional Brillouin zone of the graphene. Since electron-electron
interactions favor this type of lattice distortion, it is expected to be
enhanced, and thus its quantitative implications might be measurable in
graphene.
\end{abstract}

\maketitle

\affiliation{Department of Physics, Faculty of Sciences, Ankara University, 06100\\
Tando\u{g}an, Ankara, Turkey}

\section{Introduction}

Since the discovery of graphene \cite{science306,pnas} and its
unconventional physical properties, the investigation of electronic
properties of both graphene and graphene based nanostructures have become
one of the active areas in condensed matter physics experimentally as well
as theoretically,in past few years. Theoretically, in the low-energy limit
charge carries of graphene have linear dispersion relation around so-called
Dirac points \cite{semenoff} having Fermi velocity\cite{wallace} $%
v_{F}\simeq 10^{6}\ \mathrm{m/s}$, and Dirac-Weyl equation can be safely
used within the framework of continuum description of the electronic band
structure of the graphene \cite{nature438}.

It is also well-known that both in-plane and out-of-plane phonon modes play
an important role in charge carriers dynamics of the graphene\cite%
{prl95,prb73,nature6,prl98-2007,prb76,prb78,jphys20,stauber,prb82,prl99,samsonidze,prb76(2),prb77,prb78(2),prl100,prl103,prb81,prb82(1),prb82(3),jphys24,arxiv1,kandemir,Park2007,badalyan,krstajic,badalyan2}%
. On the one hand the Fermi velocity is reduced by interaction of charge
carriers with doubly degenerate in-plane $\mathrm{E}_{2g}$ phonon \cite%
{kandemir,Park2007}. On the other hand, though the electron-highest
frequency zone-boundary phonon interaction, i.e., Kekul\'{e}-type distortion
of the graphene lattice is one of the possible mechanisms among the gap
generations, except that the work of Ref.\onlinecite{samsonidze}, there are
no theoretical works about its influence on the graphene band dispersion.
This first theoretical prediction of dynamical mini band gap formation in
graphene due to the highest frequency phonon mode with $\mathrm{A}_{1g}$%
-symmetry is reported by Samsonidze et al. \cite{samsonidze}. They showed
that, based on a simple tight-binding model at room temperature, such as
electron-phonon coupling mechanism induces a mini gap around $10\ \mathrm{meV%
}$, and it is also responsible for the Kohn anomalies \cite{prl93,prb75} in
graphene. The Kekul\'{e} structure consists of a network of hexagons with
the alternating short and long bonds like in the classical benzene molecule.
This pattern was studied for $\mathrm{1D}$ simple model, finite size carbon
nanotubes\cite{prb45,prb47}. Investigation of the gap formation, in
particular, its control, in both graphene and graphene based nanostructures
is itself one of the hot topics of the current research in graphene, and
such a gap generation can be created by strain\cite%
{physica,acsnano1,acsnano2} or by substrate induced effects\cite%
{prb76,nature2007,prb78(3)}.

In this paper, to investigate the effect of interaction of graphene charge
carriers with the highest frequency optical phonon mode of $\mathrm{A}_{1g}$
symmetry near the zone boundary $\mathrm{K}$ ($\mathrm{K}^{\prime }$), we
performed an analytical study based on Lee-Low and Pines (LLP) theory\cite%
{pines}. The carrier-phonon interaction is described through a Kekul\'{e}%
-type distortion giving rise to inter-valley scattering between $\mathrm{K}$
and $\mathrm{K}^{\prime }$ points in graphene\cite%
{japan4,japan5,japan6,japan3,prb84}. Based on this interaction, we first
construct a Fr\"{o}lich type Dirac-Weyl Hamiltonian which is nondiagonal in
phonon creation and annihilation operators. Secondly, we present a simple
analytical model to diagonalize it by just introducing two successive
unitary transformations. Finally, we show that the interaction of charge
carriers with highest zone boundary phonon mode opens a mini band gap at the
corners of the Brillouin zone.

\section{Theory}

In the long-wave length regime, the Hamiltonian of the graphene electron
(hole) interacting with A$_{1g}$-phonon mode can be written as

\begin{equation}
\mathcal{H}=\mathcal{H}_{0}+\sum\limits_{\mu \neq \nu }\sum\limits_{%
\boldsymbol{q}}\hbar \omega _{\mu }(\boldsymbol{q})b_{\mu ,\boldsymbol{q}%
}^{\dag }b_{\mu ,\boldsymbol{q}}+\mathcal{H}_{e-p}  \label{1}
\end{equation}%
where $\mathcal{H}_{0}=v_{F}\boldsymbol{\alpha }\cdot \boldsymbol{p}$ is the
unperturbed part, whose spectrum describes cone like behavior around the
Dirac points with eigenvalues $\epsilon _{k\lambda }=\lambda v_{F}k$. $%
\lambda $ is the chirality index, and takes $-1$($+1$) values corresponding
to valence (conduction) bands in pristine graphene. These two bands touch
each other at the corners of the Brillouin zone, i.e., at the well-known $%
\mathrm{K}$ and $\mathrm{K}^{\prime }$ points whose coordinates are given by 
$\mathbf{K=}\left( 2\pi /a\right) \left( 1/3,1/\sqrt{3}\right) $ and, $%
\mathbf{K}^{\prime }\mathbf{=}\left( 2\pi /a\right) \left( 2/3,0\right) $,
respectively. We have labeled these points in Eq.(\ref{1}) by the valley
index $\mu $. Here, $\boldsymbol{\alpha }$ are the four component Dirac
matrices, and $a$ is the equilibrium bond length, i.e., $1.42\ {\mathring{A}}
$. Thus, the corresponding eigenfunctions of the unperturbed part $\mathcal{H%
}_{0}$ can easily be constructed in terms of four component pseudospinors

\begin{eqnarray}
\left\langle r\right\vert \mathrm{K}\lambda \boldsymbol{k}\rangle &=&\frac{%
\exp (i\boldsymbol{k}\cdot \boldsymbol{r})}{\sqrt{2}L}\left( 
\begin{array}{c}
\lambda \\ 
e^{i\theta \left( \boldsymbol{k}\right) } \\ 
0 \\ 
0%
\end{array}%
\right)  \notag \\
\text{ \ \ \ \ }\left\langle r\right\vert \mathrm{K}^{\prime }\lambda 
\boldsymbol{k}\rangle &=&\frac{\exp (i\boldsymbol{k}\cdot \boldsymbol{r})}{%
\sqrt{2}L}\left( 
\begin{array}{c}
0 \\ 
0 \\ 
e^{i\theta \left( \boldsymbol{k}\right) } \\ 
\lambda%
\end{array}%
\right) ,  \label{2}
\end{eqnarray}%
where $L^{2}$ is the total area of the system. In Eq.(\ref{1}), the last
term represents the electron- phonon couplings \cite{japan3}, and is given by

\begin{equation}
\mathcal{H}_{e-p}=2\frac{\beta _{\mathrm{K}}\gamma }{a^{2}}\left( 
\begin{array}{cc}
0 & \omega ^{-1}\Delta _{\mathrm{K}^{\prime }}(\boldsymbol{r})\mathbf{\sigma 
}_{y} \\ 
\omega \Delta _{\mathrm{K}}(\boldsymbol{r})\mathbf{\sigma }_{y} & 0%
\end{array}%
\right)  \label{3}
\end{equation}%
where $\beta _{\mathrm{K}}=-d\ln J_{0}/d\ln a$, $\gamma =\left( 3a/2\right)
J_{0}$, $\omega =\exp \left( 2\pi i/3\right) $, $J_{0}$ is the resonance
integral between nearest neighbor carbon atoms which is of order of $2.77\ 
\mathrm{eV}$, $\mathbf{\sigma }_{y}$ is the $2\times 2$ Pauli matrix. In Eq.(%
\ref{3}), the amplitude of distortions at $\mathrm{K}\ $\ and $\mathrm{K}%
^{\prime }$ points are defined by

\begin{eqnarray}
\Delta _{\mathrm{K}}\left( \boldsymbol{r}\right) &=&\sum_{\boldsymbol{q}}%
\sqrt{\frac{\hbar }{2NM_{C}\omega _{\text{\textrm{K}}}(\boldsymbol{q})}}%
\left( b_{\mathrm{K},\boldsymbol{q}}+b_{\mathrm{K}^{\prime },-\boldsymbol{q}%
}^{\dag }\right) e^{i\boldsymbol{q}\cdot \boldsymbol{r}}  \notag \\
\Delta _{\mathrm{K}^{\prime }}\left( \boldsymbol{r}\right) &=&\sum_{%
\boldsymbol{q}}\sqrt{\frac{\hbar }{2NM_{C}\omega _{\text{\textrm{K}}}(%
\boldsymbol{q})}}\left( b_{\mathrm{K}^{\prime },\boldsymbol{q}}+b_{\mathrm{K}%
,-\boldsymbol{q}}^{\dag }\right) e^{i\boldsymbol{q}\cdot \boldsymbol{r}},
\label{4}
\end{eqnarray}%
respectively. $N$ is the number of unit cells, $M_{C}$ is the mass of a
carbon atom. In Eq.(\ref{4}), $b_{\mathrm{K},\boldsymbol{q}}$ ($b_{\mathrm{K}%
^{\prime },\boldsymbol{q}}$) and $b_{\mathrm{K},\boldsymbol{q}}^{\dag }$ ($%
b_{\mathrm{K}^{\prime },\boldsymbol{q}}^{\dag }$) are the phonon creation
and annihilation operators at points $\mathrm{K}$ ($\mathrm{K}^{\prime }$)
with phonon wave vector $\boldsymbol{q}$ and frequency $\omega _{\mathrm{K}}(%
\boldsymbol{q})$. The corresponding highest zone-boundary phonon energy is $%
\hbar \omega _{\mathrm{K}}(0)=161.2$ meV. Therefore, the electron-phonon
interaction Hamiltonian given by Eq.(\ref{3}) can be conveniently rewritten
in the following form:

\begin{equation}
\mathcal{H}_{e-p}=-\sum\limits_{\mu \neq \nu }\sum\limits_{\boldsymbol{q}}%
\left[ \widetilde{\mathrm{M}}_{\mu \nu }b_{\mu ,\boldsymbol{q}}e^{i%
\boldsymbol{q\cdot r}}+\text{h.c.}\right] .  \label{5}
\end{equation}%
We have defined $\widetilde{\mathrm{M}}_{\mu \nu }$ as $\mathrm{M}_{0}%
\mathrm{M}_{\mu \nu }$ such that 
\begin{eqnarray*}
\mathrm{M}_{\mathrm{KK}^{\prime }} &=&\frac{\omega }{\sqrt{N}}\left( 
\begin{array}{cl}
\mathbf{0} & \mathbf{0} \\ 
\mathbf{\sigma }_{y} & \mathbf{0}%
\end{array}%
\right) , \\
\mathrm{M}_{\mathrm{K}^{\prime }\mathrm{K}} &=&\frac{\omega ^{-1}}{\sqrt{N}}%
\left( 
\begin{array}{lc}
\mathbf{0} & \mathbf{\sigma }_{y} \\ 
\mathbf{0} & \mathbf{0}%
\end{array}%
\right) ,
\end{eqnarray*}%
together with \textrm{M}$_{0}=3a_{0}q_{0}J_{0}$. Here, $a_{0}=\left( \hbar
/2M_{C}\omega _{\mathrm{K}}(0)\right) ^{1/2}$, and $q_{0}=\left( \partial
J_{0}/\partial a\right) /J_{0}$ is predicted \cite{prb48,Pietronero1980}
around $2.0$ ${\mathring{A}}^{-1}$and $2.5$ ${\mathring{A}}^{-1}$.

To diagonalize the phonon subsystem of Eq.(\ref{1}) through Eq.(\ref{5}) we
employ a unitary transformation scheme within the LLP theory. This includes
two successive transformations each of which eliminates the electron
coordinates from Eq.(\ref{1}), and shifts phonon coordinates by an amount of
the interaction strength, respectively. To do this we follow the method
developed for the investigation of the interaction of electron (hole) with
doubly degenerate optical phonon modes of $\mathrm{E}_{2g}$ symmetry near
the zone center \cite{kandemir}, wherein an ansatz was made so as to take
into account chiral nature of the pristine graphene due to its gapless
electronic band structure. However, besides the chiral nature of the
problem, one must also considered that the zone boundary phonon gives rise
to inter-valley scattering between $\mathrm{K}$ and $\mathrm{K}^{\prime }$.
Therefore, to be compatible with these properties of the problem, we make an
ansatz for the ground-state of the whole system. 
\begin{equation}
\mid \boldsymbol{\Phi }\rangle =\sum_{\mu ^{\prime }\neq \nu ^{\prime
}}\sum\limits_{\lambda ^{\prime }}\alpha _{\pm }^{\mu ^{\prime }\lambda
^{\prime }}\mid \mu ^{\prime }\lambda ^{\prime }\boldsymbol{k}\rangle
\otimes U_{1}U_{2}\mid \boldsymbol{0}\rangle _{\mathrm{ph}}  \label{6}
\end{equation}%
such that $\mathcal{H}\mid \boldsymbol{\Phi }\rangle =E_{\pm }\mid 
\boldsymbol{\Phi }\rangle $. Here, $\mid \boldsymbol{0}\rangle _{\mathrm{ph}%
} $ stands for the phonon vacuum, and $\alpha _{\pm }^{\mu ^{\prime }\lambda
^{\prime }}\mid \mu ^{\prime }\lambda \boldsymbol{k}\rangle $ corresponds to
electronic state vector defined through the appropriate fractional
amplitudes, $\alpha _{\pm }^{\mu ^{\prime }\lambda }$, due to the fact that
total wave function of the system must be the linear combination of $\mid
\mu ^{\prime }+\boldsymbol{k}\rangle $ and $\mid \mu ^{\prime }-\boldsymbol{k%
}\rangle $, respectively.

On the one hand, the first unitary transformation

\begin{equation}
U_{1}=\exp \left[ -i\boldsymbol{r\cdot }\sum\limits_{\boldsymbol{q}}%
\boldsymbol{q}b_{\mu ,\boldsymbol{q}}^{\dag }b_{\mu ,\boldsymbol{q}}\right]
\label{7}
\end{equation}%
eliminates electron coordinates from Eq.(\ref{1}), since the transformed
operators are given by the relations, $\widetilde{b}_{\mu ,\boldsymbol{q}%
}=b_{\mu ,\boldsymbol{q}}\exp \left[ -i\boldsymbol{q\cdot r}\right] $ and $%
\widetilde{\boldsymbol{p}}=\boldsymbol{p}-\sum\limits_{\boldsymbol{q}%
}\sum\limits_{\mu \neq \nu }\hbar \boldsymbol{q}b_{\mu ,\boldsymbol{q}%
}^{\dag }b_{\mu ,\boldsymbol{q}}$. Therefore, the transformed Hamiltonian
takes the form,

\begin{eqnarray}
\bigskip \widetilde{\mathcal{H}} &=&v_{F}\boldsymbol{\alpha }\mathbf{\cdot }%
\left( \boldsymbol{p}-\hbar \sum\limits_{\boldsymbol{q}}\sum\limits_{\mu
\neq \nu }\boldsymbol{q}b_{\mu ,\boldsymbol{q}}^{\dag }b_{\mu ,\boldsymbol{q}%
}\right)  \notag \\
&&+\sum\limits_{\boldsymbol{q}}\sum\limits_{\mu \neq \nu }\hbar \omega _{%
\mathrm{K}}b_{\mu ,\boldsymbol{q}}^{\dag }b_{\mu ,\boldsymbol{q}%
}-\sum\limits_{\boldsymbol{q}}\sum\limits_{\mu \neq \nu }\left( \widetilde{%
\text{M}}_{\mu \nu }b_{\mu ,\boldsymbol{q}}+\text{h.c.}\right) .  \notag \\
&&  \label{8}
\end{eqnarray}%
On the other hand, second unitary transformation

\begin{equation}
U_{2}=\exp \left[ \sum\limits_{\boldsymbol{q}}\widetilde{\mathrm{M}}%
_{0}\langle \mu ^{\prime }\lambda ^{^{\prime }}\boldsymbol{k}\mid \mathrm{M}%
_{\mu \nu }^{\dag }\mid \nu ^{\prime }\lambda \boldsymbol{k}\rangle b_{\mu 
\boldsymbol{,q}}^{\dag }-\text{h.c}.\right]  \label{9}
\end{equation}%
is the well-known displaced oscillator transformation which shifts phonon
coordinates by an amount of the interaction amplitude, $\widetilde{\text{M}}%
_{0}=$M$_{0}/\hbar \omega _{\mathrm{K}}(0)$. It just shifts the phonon
coordinates, since it generates the coherent states for the phonon subsystem
such that optical phonon operators transform according to the rule $%
\widetilde{b}_{\mu ,\boldsymbol{q}}=b_{\mu ,\boldsymbol{q}}+\widetilde{%
\mathrm{M}}_{0}\langle \mu ^{\prime }\lambda ^{^{\prime }}\boldsymbol{k}\mid 
\mathrm{M}_{\mu \nu }^{\dag }\mid \nu ^{\prime }\lambda \boldsymbol{k}%
\rangle $. As a result, under the transformation $U_{2}$, Eq.(\ref{8}) can
then be written as $\widetilde{\mathcal{H}}=\mathcal{H}^{0}+\mathcal{H}_{1}$%
, where $\mathcal{H}^{0}$ and $\mathcal{H}_{1}$ are given by 
\begin{widetext}
\begin{eqnarray}
\mathcal{H}^{0} &=&v_{F}\boldsymbol{\alpha }\mathbf{\cdot }\left( 
\boldsymbol{p}-\hbar \sum\limits_{\boldsymbol{q}}\sum\limits_{\mu \neq \nu
}\left\vert \widetilde{\text{M}}_{0}\right\vert ^{2}\boldsymbol{q}\left\vert
\langle \delta \lambda ^{^{\prime }}\boldsymbol{k}\mid \text{M}_{\mu \nu
}^{\dag }\mid \zeta \lambda \boldsymbol{k}\rangle \right\vert ^{2}\right)
+\sum\limits_{\boldsymbol{q}}\sum\limits_{\mu \neq \nu }\left\vert 
\widetilde{\text{M}}_{0}\right\vert ^{2}\hbar \omega _{\text{\textrm{K}}%
}\left\vert \langle \delta \lambda ^{^{\prime }}\boldsymbol{k}\mid \text{M}%
_{\mu \nu }^{\dag }\mid \zeta \lambda \boldsymbol{k}\rangle \right\vert ^{2}
\notag \\
&&-\sum\limits_{\boldsymbol{q}}\sum\limits_{\mu \neq \nu }\left[ \left\vert 
\widetilde{\text{M}}_{0}\right\vert ^{2}\hbar \omega _{\text{\textrm{K}}}%
\text{M}_{\mu \nu }\left\vert \langle \delta \lambda ^{^{\prime }}%
\boldsymbol{k}\mid \text{M}_{\mu \nu }^{\dag }\mid \zeta \lambda \boldsymbol{%
k}\rangle \right\vert ^{2}+\text{h.c.}\right] +\sum\limits_{\boldsymbol{q}%
}\sum\limits_{\mu \neq \nu }\left[ \hbar \omega _{\mu }(\boldsymbol{q}%
\mathbf{)-}\hbar v_{F}\boldsymbol{\alpha }\cdot \boldsymbol{q}\right] b_{\mu
,\boldsymbol{q}}^{\dag }b_{\mu ,\boldsymbol{q}},  \label{10}
\end{eqnarray}%
\end{widetext}
and 
\begin{widetext}
\begin{equation}
\mathcal{H}_{1}=\sum\limits_{\boldsymbol{q}}\sum\limits_{\mu \neq \nu
}\left\{ \left[ \mathrm{M}_{\mu \nu }+\left[ \hbar \omega _{\mu }(\mathbf{{q}%
)-}\hbar v_{F}\boldsymbol{\alpha }\mathbf{\cdot {q}}\right] \widetilde{%
\mathrm{M}}_{0}\langle \mu ^{\prime }\lambda ^{^{\prime }}\boldsymbol{k}\mid 
\mathrm{M}_{\mu \nu }\mid \nu ^{\prime }\lambda \boldsymbol{k}\rangle \right]
b_{\mu ,\boldsymbol{q}}+\mathrm{H.c.}\right\} ,  \label{11}
\end{equation}%
\end{widetext}
respectively. Therefore, one applies the phonon vacuum to the sum of $\ $Eq.(%
\ref{10}) and Eq.(\ref{11}), only the contribution comes from the
diagonalized part , i.e. from $\mathcal{H}^{0}$. By using the ansatz given
by Eq.(\ref{6}) , one first applies $\ $Eq.(\ref{10}) to the term $\alpha
_{\pm }^{\mu ^{\prime }\lambda ^{\prime }}\mid \mu ^{\prime }\lambda
^{\prime }\boldsymbol{k}\rangle $, and then sums over $\mathbf{\lambda
^{\prime }}$\ \ to construct the eigenvalue equation $\mathcal{H}\mid 
\boldsymbol{\Phi }\rangle ^{\mu \nu \lambda }=E_{\pm }\mid \boldsymbol{\Phi }%
\rangle ^{\mu \nu \lambda }$. Finally, by taking inner products to compare
the related coefficients of the states$\mid \mu ^{\prime }\lambda ^{\prime }%
\boldsymbol{k}\rangle $ we arrive four simultaneous equations for $\alpha
_{\pm }^{\mathbf{K}+}$ , $\alpha _{\pm }^{\mathbf{K}-}$, $\alpha _{\pm }^{%
\mathbf{K}^{\prime }+}$ and $\alpha _{\pm }^{\mathbf{K}^{\prime }-}$which
can be rewritten in the following matrix equation: 
\begin{widetext}
\begin{equation}
\begin{bmatrix}
E_{\pm }+\mathbf{\Sigma }_{++}^{(0)\mathrm{KK}} & \mathbf{\Sigma }_{--}^{(1)%
\text{\textrm{KK}}} & \mathbf{\Sigma }_{++}^{(2)\text{\textrm{KK}}^{\prime }}
& \mathbf{\Sigma }_{+-}^{(2)\text{\textrm{KK}}^{\prime }} \\ 
\mathbf{\Sigma }_{++}^{(1)\mathrm{KK}} & E_{\pm }+\mathbf{\Sigma }_{--}^{(0)%
\text{\textrm{KK}}} & \mathbf{\Sigma }_{-+}^{(2)\text{\textrm{KK}}^{\prime }}
& \mathbf{\Sigma }_{++}^{(2)\text{\textrm{KK}}^{\prime }} \\ 
\mathbf{\Sigma }_{++}^{(2)\text{\textrm{K}}^{\prime }\text{\textrm{K}}} & 
\mathbf{\Sigma }_{+-}^{(2)\text{\textrm{K}}^{\prime }\text{\textrm{K}}} & 
E_{\pm }+\mathbf{\Sigma }_{++}^{(0)\text{\textrm{K}}^{\prime }\text{\textrm{K%
}}^{\prime }} & \mathbf{\Sigma }_{--}^{(1)\text{\textrm{K}}^{\prime }\text{%
\textrm{K}}^{\prime }} \\ 
\mathbf{\Sigma }_{-+}^{(2)\text{\textrm{K}}^{\prime }\text{\textrm{K}}} & 
\mathbf{\Sigma }_{--}^{(2)\text{\textrm{K}}^{\prime }\text{\textrm{K}}} & 
\mathbf{\Sigma }_{++}^{(1)\text{\textrm{K}}^{\prime }\text{\textrm{K}}%
^{\prime }} & E_{\pm }+\mathbf{\Sigma }_{--}^{(0)\text{\textrm{K}}^{\prime }%
\text{\textrm{K}}^{\prime }}%
\end{bmatrix}%
\left[ 
\begin{array}{c}
\alpha _{\pm _{{}}}^{\text{\textrm{K}}_{{}}^{{}}+} \\ 
\alpha _{\pm _{{}}}^{\text{\textrm{K}}_{{}}^{{}}-} \\ 
\alpha _{\pm _{{}}}^{\text{\textrm{K}}^{\prime }+} \\ 
\alpha _{\pm _{{}}}^{\text{\textrm{K}}^{\prime }-}%
\end{array}%
\right] =0  \label{12a}
\end{equation}%
\end{widetext}
with elements 
\begin{eqnarray}
\mathbf{\Sigma }_{\mp \mp }^{(0)\text{\textrm{KK}}} &=&\pm \hbar
v_{F}k+\sum\limits_{\boldsymbol{q}}\sum\limits_{\boldsymbol{\mu \neq \nu }%
}\left\vert \widetilde{\text{M}}_{0}\right\vert ^{2}\Delta _{\mp \mp }^{%
\text{\textrm{KK}}}\left[ \frac{1}{2}\hbar v_{F}q\Theta _{\mp \mp }+\hbar
\omega _{\text{\textrm{K}}}\right]  \notag \\
\mathbf{\Sigma }_{\mp \mp }^{(1)\text{\textrm{KK}}} &=&\sum\limits_{%
\boldsymbol{q}}\sum\limits_{\boldsymbol{\mu \neq \nu }}\left\vert \widetilde{%
\text{M}}_{0}\right\vert ^{2}\Delta _{\mp \mp }^{\text{\textrm{KK}}}\left[
\hbar v_{F}q\Delta _{\mp \mp }^{\text{\textrm{KK}}}\Theta _{\mp \pm }+2\hbar
\omega _{\mathrm{K}}\Delta _{\mp \pm }^{\text{\textrm{KK}}}\right]  \notag \\
\mathbf{\Sigma }_{\lambda \lambda ^{\prime }}^{(2)\text{\textrm{KK}}^{\prime
}} &=&2\hbar \omega _{\text{\textrm{K}}}\sum\limits_{\boldsymbol{q}%
}\sum\limits_{\boldsymbol{\mu \neq \nu }}\left\vert \widetilde{\text{M}}%
_{0}\right\vert ^{2}\Delta _{\lambda \lambda ^{\prime }}^{\text{\textrm{KK}}%
^{\prime }}.  \label{12}
\end{eqnarray}%
where we have defined $\Theta _{\lambda \lambda ^{\prime }}$ as 
\begin{equation}
\Theta _{\lambda \lambda ^{\prime }}=\frac{1}{2}\left[ \left( s+s^{\prime
}\right) \cos (\theta -\varphi )-\left( s-s^{\prime }\right) \sin (\theta
-\varphi )\right]  \label{13}
\end{equation}%
together with the matrix elements 
\begin{equation}
\Delta _{\lambda \lambda ^{\prime }}^{\zeta \zeta ^{\prime }}=\text{\ \ \ }%
\left\vert \langle \zeta ^{\prime }\lambda ^{^{\prime }}\boldsymbol{k}\mid 
\mathrm{M}_{\mu \nu }\mid \zeta \lambda \boldsymbol{k}\rangle \right\vert
^{2}  \label{14}
\end{equation}%
which are equal to unity for $\zeta \neq \zeta ^{\prime }$ and $\lambda \neq
\lambda ^{\prime }$, otherwise zero. \ This shows that only inter-valley
scattering having different chiralities are allowed due to the conservation
of the chiral symmetry . In Eq.(\ref{13}), $\theta (\phi )$ are the
azimuthal angle of the momentum $\boldsymbol{k}$($\boldsymbol{q}$). After
converting the sums in Eq.(\ref{12}) into integrals over $\boldsymbol{q}$,
i.e., $\sum_{\boldsymbol{q}}\rightarrow \left( S/4\pi ^{2}\right) \int d^{2}%
\boldsymbol{q}$, where $S=NS_{0}$ is the area of the system, and the area of
the unit cell is $S_{0}=3\sqrt{3}a^{2}/2$, it is easy to see that, except
for $\mathbf{\Sigma }_{\mp \pm }^{(2)\text{\textrm{KK}}^{\prime }}$and $%
\mathbf{\Sigma }_{\mp \pm }^{(2)\text{\textrm{K}}^{\prime }\text{\textrm{K}}%
} $\ terms, all the terms with $\mathbf{\Sigma }_{\mp \mp }^{(0)\text{%
\textrm{KK}}}$, $\mathbf{\Sigma }_{\mp \mp }^{(1)\text{\textrm{KK}}}$ and
their corresponding \textrm{K}$^{\prime }$ partners vanish. \ The
non-vanishing terms \ can easily be calculated as

\begin{equation}
\mathbf{\Sigma }_{\mp \pm }^{(2)\text{\textrm{KK}}^{\prime }}=\mathbf{\Sigma 
}_{\mp \pm }^{(2)\text{\textrm{K}}^{\prime }\text{\textrm{K}}}=\frac{3\sqrt{3%
}}{\pi }J_{0}\alpha _{0}\overline{q}_{c}^{2}  \label{15}
\end{equation}%
where we have defined that $\alpha _{0}=\left\vert \mathrm{M}_{0}\right\vert
^{2}/4J_{0}\hbar \omega _{\text{\textrm{K}}}(0)$ which takes values $0.305$
or $0.477$ depending on whether $q_{0}$ is $2.0$ ${\mathring{A}}^{-1}$ or $%
2.5$ ${\mathring{A}}^{-1}$, respectively\cite{prb48,Pietronero1980}. We must
introduce an upper cut-off frequency $\overline{q}_{c}=q_{c}a$, while taking
the integrals in Eq.(\ref{12}), since they diverge at upper limit of the
integrations. By solving the determinant of the matrix in Eq.(\ref{12a}) the
eigenvalues $E_{\pm }$ can be solved analytically in the following form:

\begin{equation}
E_{\pm }=\pm \left[ \left( \hbar v_{F}\boldsymbol{k}\right) ^{2}+\left( 
\mathbf{\Sigma }_{\mp \pm }^{(2)\text{\textrm{KK}}^{\prime }}\right) ^{2}%
\right] ^{1/2},  \label{16}
\end{equation}%
which is modified electronic band dispersion of the pristine graphene by the
gap due to the Kekul\'{e}-type distortion of the lattice. As is seen from
Eq.(\ref{16}), \ Kekul\'{e}-type distortion preserves the chirality of the
sublattice, i.e., the valley degeneracy is not lifted. Since the
calculations we have done are restricted by the energy scale near to the
phonon resonance, i.e., $\omega _{\text{\textrm{K}}}$, we can choose $%
q_{c}=\omega _{\text{\textrm{K}}}/v_{F}=0.027$ ${\mathring{A}}^{-1}$ such
that $\overline{q}_{c}=q_{c}a=0.039$. This suggest that the magnitude of the
half-band gap is of order of $2.12$ meV ($q_{0}=2.0$ ${\mathring{A}}^{-1}$)
or $3.34$ meV ($q_{0}=2.5$ ${\mathring{A}}^{-1}$), so that the induced gap,
of $4.24$ meV or $6.68$ meV. This is smaller than that those previously
found in the literature\cite{samsonidze}, where a mini band gap occurs $10$
meV in their room temperature calculations.

\section{ Conclusion}

In conclusion, we present, to the best our knowledge, the first theoretical
justification of a mini band gap formation in pristine graphene at absolute
zero temperature, due to the interaction of electron (hole) with highest
frequency optical phonon mode with A$_{1g}$ symmetry near the zone boundary
yielding Kekul\'{e} type distortion. We have shown that such an interaction
opens a gap without breaking the chiral symmetry of the lattice. Although,
we have performed our calculations at absolute zero, it is compatible with
those found at room temperatures for both graphene and zig-zag single walled
carbon nanotubes \cite{prb75}. In the graphene literature, there are only
two theoretical works\cite{prb84,prb82(4)} and no experimental work devoted
to the investigation of combined effects of electron-electron interaction
and electron-\textrm{A}$_{1g}$ phonon interaction. On the one hand, since
the electron repulsion breaks the sublattice symmetry whereas the Kekul\'{e}
patterned lattice distortion restores it, it is predicted in Ref. %
\onlinecite{prb84} that , taking such competing interactions,
superconducting order maybe have to be considered. On the other hand,
Giuliani et al.\cite{prb82(4)} predicted that the electron repulsion
enhances dramatically, the gaps due to the Kekul\'{e} distortion. We believe
that the results of this paper may be useful for understanding the role of
electron-electron interactions in graphene.

\begin{acknowledgements}
We thank Professor T. Altanhan for valuable discussions. The authors
gratefully acknowledge the support from the Research Projects of Ankara
University under grant No. 12B4240009.
\end{acknowledgements}


\begin{thebibliography}{99}
\bibitem{science306} K. S. Novoselov, A. K. Geim, S. V. Morozov, D. Jiang,
Y. Zhang, S. V. Dubonos, I. V. Grigorieva and A. A. Firsov, Science \textbf{%
306}, 666 (2004).

\bibitem{pnas} K. S. Novoselov, D. Jiang, F. Schedin, T. J. Booth, V. V.
Khotkevich, S.V. Morozov, and A. K. Geim, Proc. Nat. Acad. Sci. USA \textbf{%
102}, 10451 (2005).

\bibitem{semenoff} G. W. Semenoff, Phys. Rev. Lett. \textbf{53}, 2449 (1984).

\bibitem{wallace} P.R. Wallace, Phys. Rev. \textbf{71},622 (1947).

\bibitem{nature438} K. S. Novoselov, A. K. Geim, S. V. Morozov, D. Jiang, M.
I. Katsnelson, I. V. Grigorieva, S. V. Dubonos, and A. A. Firsov, Nature 
\textbf{438}, 197-200 (2005).

\bibitem{prl95} Michele Lazzeri, S. Piscanec, Francesco Mauri, A. C.
Ferrari, and J. Robertson, Phys. Rev. Lett. \textbf{95} 236802, (2005).

\bibitem{prb73} Michele Lazzeri, S. Piscanec, Francesco Mauri, A. C.
Ferrari, and J. Robertson, Phys. Rev. B \textbf{73}, 155426 (2006).

\bibitem{nature6} Simone Pisana, Michele Lazzeri, Cinzia Casiraghi, Kostya
S. Novoselov, A. K. Geim, Andrea C. Ferrari and Francesco Mauri, Nature
Materials \textbf{3},198 (2007).

\bibitem{prl98-2007} Jun Yan,Yuanbo Zhang, Philip Kim, and Aron Pinczuk,
Phys. Rev.Lett. \textbf{98}, 166802 (2007).

\bibitem{prb76} Matteo Calandra and Francesco Mauri, Phys Rev. B \textbf{76}%
,205411 (2007).

\bibitem{prb78} D. M. Basko, Phys. Rev. B \textbf{78},125418 (2008).

\bibitem{jphys20} T Stauber and N M R Peres, J. Phys.:Condensed Matter 
\textbf{20},055002, (2008).

\bibitem{stauber} T. Stauber, N. M. R. Peres, and A. H. Castro Neto, Phys.
Rev. B \textbf{78}, 085418 (2008).

\bibitem{prb82} Vladimir M. Stojanovi\'{c},Nenad Vukmirovi\'{c}, C. Bruder,
Phys. Rev. B \textbf{82}, 165410, (2010).

\bibitem{prl99} M. O. Goerbig, J.-N. Fuchs, K. Kechedzhi and Vladimir I.
Fal'ko, Phys. Rev. Lett. \textbf{99}, 087402 (2007).

\bibitem{samsonidze} Ge. G. Samsonidze, E. B. Barros, R. Saito, J. Jiang, G.
Dresselhaus, and M. S. Dresselhaus, Phys. Rev. B \textbf{75}, 155420 (2007).

\bibitem{prb76(2)} D. M. Basko, Phys. Rev. B \textbf{76}, 081405 (R) (2007).

\bibitem{prb77} D.M. Basko, I.L. Aleiner, Phys. Rev. B \textbf{77} (2008)
041409(R).

\bibitem{prb78(2)} Michele Lazzeri, Claudio Attaccalite, Ludger Wirtz, and
Francesco Mauri, Phys. Rev. B \textbf{78}, 081406 (R), (2008).

\bibitem{prl100} Eros Mariani and Felix von Oppen, Phys. Rev. Lett. \textbf{%
100}, 076801 (2008).

\bibitem{prl103} C. Faugeras, M. Amado, P. Kossacki, M. Orlita, M. Sprinkle,
C. Berger, W. A. de Heer, and M. Potemski, Phys. Rev. Lett. \textbf{103},
186803, (2009).

\bibitem{prb81} J. P. Carbotte, E. J. Nicol, and S. G. Sharapov, Phys. Rev.
B \textbf{81},045419, (2010).

\bibitem{prb82(1)} Eros Mariani and Felix von Oppen, Phys. Rev. B \textbf{82}%
, 195403, (2010).

\bibitem{prb82(3)} E. H. Hwang, Rajdeep Sensarma, and S. Das Sarma, Phys.
Rev. B \textbf{82}, 195406, (2010).

\bibitem{jphys24} Wei-Ping Li, Zi-Wu Wang, Ji-Wen Yin and Yi-Fu Yu, J.
Phys.: Condensed Matt. \textbf{24}, 135301, (2012).

\bibitem{arxiv1} P. T. Araujo, D. L. Mafra, K. Sato, R. Saito, J. Kong and
M. S. Dresselhaus, arXiv:1203.0547v1 (2012).

\bibitem{kandemir} B. S. Kandemir, J. Phys.: Condensed Matt. \textbf{25},
025302, (2013)

\bibitem{Park2007} Cheol-Hwan Park, Feliciano Giustino, Marvin L Cohen , and
Steven G Louie Phys.Rev.Lett\textit{.} \textbf{99}, 086804 (2007)

\bibitem{badalyan} S. M. Badalyan and F. M. Peeters, Phys. Rev. B \textbf{85}%
, 205453 (2012).

\bibitem{krstajic} P.M. Krstajic and F. M. Peeters, Phys. Rev. B \textbf{85}%
, 205454 (2012).

\bibitem{badalyan2} J. Zhu, S. M. Badalyan, and F. M. Peeters, Phys. Rev.
Lett. \textbf{109}, 256602 (2012).

\bibitem{prl93} S. Piscanec, M. Lazzeri, Francesco Mauri, A. C. Ferrari, and
J. Robertson, Phys Rev. Lett. \textbf{93}, 185503 (2004).

\bibitem{prb75} Ge. G. Samsonidze, E. B. Barros, R. Saito, J. Jiang, G.
Dresselhaus, and M. S. Dresselhaus, Phys. Rev. B \textbf{75}, 155420 (2007).

\bibitem{prb45} Kikuo Harigaya, Phys. Rev. B, \textbf{45}, 12071 (1992).

\bibitem{prb47} Kikuo Harigaya and Mitsutaka Fujita, Phys. Rev. B \textbf{47}%
, 16563 (1993).

\bibitem{physica} M. Farjam and H. Rafii-Tabar, Physica E \textbf{42},
2109-2114 (2010).

\bibitem{acsnano1} Sung-Hoon Lee, Hyun-Jong Chung, Jinseong Heo, Heejun
Yang, Jaikwang Shin, U-\i n Chung, and Sunae Seo, ACS Nano, \textbf{5},
2964-2969 (2011).

\bibitem{acsnano2} Zhen Hua Ni, Ting Yu, Yun Hao Lu, Ying Ying Wang, Yuan
Ping Feng, and Ze Xiang Shen, ACS Nano, \textbf{2}, 2301-2305 (2008).

\bibitem{nature2007} S.Y. Zhou, G. H. Gweon, A.V. Federov, P. NFirst, W. A.
De Heer, D.-H. Lee, F. Guinea, A. H. Castro Neto and A. Lanzara, Nature, 
\textbf{6}, 770 (2007).

\bibitem{prb78(3)} S. Y. Zhou, D. A. Siegel, A. V. Fedorov and A. Lanzara,
Phys. Rev. B. \textbf{78}, 193404 (2008).

\bibitem{pines} T.D. Lee, F.E. Low, and D. Pines, Phys. Rev. \textbf{90},
297-302 (1953).

\bibitem{japan4} Nguyen Ai Viet, Hiroshi Ajiki and Tsuneya Ando, J. Phys.
Soc. Jpn. \textbf{6}3, 3036-3047 (1994).

\bibitem{japan5} Hiroshi Ajiki and Tsuneya Ando, J. Phys. Soc. Jpn. \textbf{%
64}, 260-267 (1995).

\bibitem{japan6} Hiroshi Ajiki and Tsuneya Ando, J. Phys. Soc. Jpn. \textbf{%
65}, 2976-2986 (1996).

\bibitem{japan3} Hidekatsu Suzuura and Tsuneya Ando, J. Phys. Soc. Jpn. 
\textbf{77}, 044703 (2008).

\bibitem{prb84} Yasufumi Araki, Phys. Rev. B, \textbf{84}, 113402 (2011).

\bibitem{Pietronero1980} L. Pietronero , S. Str\"{a}ssler , and H. R. Zeller
, M. J. Rice, Phys. Rev. B, \textbf{22,} 904 (1980)

\bibitem{prb48} R. A. Jishi, M. S. Dresselhaus, and G. Dresselhaus, Phys.
Rev. B. \textbf{48}, 11385 (1993).

\bibitem{prb82(4)} Alessandro Giuliani, Vieri Mastropietro, and Marcello
Porta, Pjys. Rev. B \textbf{82}, 121418 (R) 2010.
\end{thebibliography}
\end{document}